\documentclass[twocolumn]{aastex63}

\usepackage{graphicx}
\usepackage{dcolumn}
\usepackage{bm}

\usepackage{makecell}
\usepackage{verbatim} 
\usepackage{amsmath}

\begin{document}


\title{Exploring the Effects of Active Magnetic Drag in a GCM of the Ultra-Hot Jupiter WASP-76b }

\correspondingauthor{Hayley Beltz}
 \email{hbeltz@umich.edu}

\author[0000-0002-6980-052X]{Hayley Beltz}

 \affiliation{Department of Astronomy, University of Michigan, Ann Arbor, MI 48109, USA}
\author[0000-0003-3963-9672]{Emily Rauscher}
\affiliation{Department of Astronomy, University of Michigan, Ann Arbor, MI 48109, USA}

\author[0000-0002-0786-7307]{Michael T. Roman}
\affil{University of Leicester,
School of Physics and Astronomy,
University Road, Leicester, LE1 7RH, UK}

\author[0000-0000-0000-0000]{Abigail Guilliat}
\affiliation{Department of Astronomy, University of Michigan, Ann Arbor, MI 48109, USA}

\begin{abstract}
Ultra-hot Jupiters represent an exciting avenue for testing extreme physics and observing atmospheric circulation regimes not found in our solar system. Their high temperatures result in thermally ionized particles embedded in atmospheric winds interacting with the planet's interior magnetic field by generating current and experiencing bulk Lorentz force drag. Previous treatments of magnetic drag in 3D General Circulation Models (GCMs) of ultra-hot Jupiters have mostly been uniform drag timescales applied evenly throughout the planet, which neglects the strong spatial dependence of these magnetic effects. In this work, we apply our locally calculated active magnetic drag treatment in a GCM of the planet WASP-76b. We find the effects of this treatment to be most pronounced in the planet's upper atmosphere, where strong differences between the day and night side circulation are present. These circulation effects alter the resulting phase curves by reducing the hotspot offset and increasing the day-night flux contrast. We compare our models to \textit{Spitzer} phase curves which imply a magnetic field of at least 3 G for the planet.  We additionally
\deleted{compare} \added{contrast} our results to uniform drag timescale models. This work highlights the need for more careful treatment of magnetic effects in atmospheric models of hot gas giants.  
\end{abstract}

\section{Introduction}

Gas giant planets orbiting extremely close to their host stars make excellent targets for observers due to \added{their} favorable planet-star flux ratios and offer avenues for testing prescriptions of high-temperature physical processes in theoretical models. The quintessential category of these planets is ultra-hot Jupiters (UHJs), which orbit so close to their host star that their equilibrium temperatures exceed $\sim$2200 K.  The extreme temperatures of these planets warrant a careful consideration of the relevant physical processes included in the models, as the increased irradiation results in different dominant mechanisms than those of their cooler cousins, ``normal'' hot Jupiters.  \added{While magnetic effects may begin to alter the circulation patterns of gas giant planets that reach temperatures $\approx 1500$ K \citep[][]{Menou_2012,Rogers_2014b}, it is not until planets reach an equilibrium temperature $\gtrsim 2000$ K that non-ideal MHD effects are predicted to become very strong due to the coupling between the circulation and atmospheric magnetic field, potentially resulting in effects such as hot spot reversals \citep[][]{Hindle2021b}.} \deleted{This is particularly true for the effects of magnetism, which is the main focus of this work.} Since no solar system analogue exists for these planets, intricate multi-dimensional atmospheric modeling is critical for understanding and interpreting observations of their atmospheres.

The consequences of magnetism manifest themselves on hot gas giants in a multitude of ways. The interior of the planet is expected to host a magnetic field of comparable or even greater strength than solar system Jovian planets \citep{Yadav2017}. In the planet's upper atmosphere, high energy stellar photons can photo-ionize species and drive evaporative winds, whose outflow can be shaped by the planet's magnetic field \citep{Owen2014}. Here, we focus on the atmospheric effects of thermal ionization of species near the planet's photosphere interacting with the planet's magnetic field. Due to the extreme temperatures on the daysides of these planets, species undergo thermal ionization while remaining embedded in the mostly neutral atmosphere; these ions then are advected around the planet via strong winds. The currents generated by this interaction could travel into the planet's interior and deposit heat there via Ohmic dissipation, perhaps explaining the inflated radii of many hot Jupiters \citep[][]{BatyginStevenson2010,Perna2010b,Thorngren2018}. \deleted{In addition,} The coupling between the ions and the mostly neutral winds results in a bulk Lorentz force drag, potentially reducing circulation efficiencies and increasing the day-night contrast \citep[][]{Perna2010,Menou_2012, Batygin2013,RauscherMenou2013,Rogers_2014_showman}.
Current treatment of magnetic effects in 3D atmospheric models vary in complexity, from computationally expensive non-ideal MHD simulations to the use of a universal drag timescale applied throughout the planetary atmosphere, with this range in modeling complexity serving a variety of purposes. 

\citet{Perna2010} introduced the idea of parameterizing the effects of the bulk Lorentz force on atmospheric winds with a drag timescale. The formulation of this timescale was derived from order of magnitude approximations of the terms in the non-ideal MHD induction equation in the hot Jupiter regime, \added{thus effectively operating as a ``kinematic'' MHD model}. This timescale was used in \citet{RauscherMenou2013} in their models of two different hot Jupiters. This work found that for the hotter planet modeled, (HD~209458b, $T_{\mathrm{eq}} \approx 1500$ K), the drag prescription resulted in slower wind speeds and a reduction in the strength of the equatorial jet often seen in hot Jupiter atmospheric models.  

One simpler treatment for magnetic drag is applying a uniform drag timescale to the entire modeled atmosphere. \citet{KomacekShowman2016} explored the use of various uniform values for their model of a HD~209458b-like hot Jupiter to approximate the effect of Lorentz forces. So long at the drag timescale was short enough ($ \lesssim 10^{4}$s), the zonal jet was eliminated and day-night temperature differences were large. Additionally, they noted that their drag timescale had only a secondary effect on the day-night temperature contrast when it was longer than the rotation rate of the planet. Uniform drag timescales are found in many works involving GCMs of hot (or ultra hot) Jupiters \citep{komacekShowman2017, Koll2018, Kreidberg_2018,Mansfield_2018,Arcangeli2019}.  These universal drag timescales can be converted to global magnetic field strengths with order-of-magnitude estimates, such as in \cite{Kreidberg_2018}, which estimated a magnetic field stronger than $\sim 1$ Gauss for the UHJ WASP-103b. This uniform drag treatment will also have an effect on the predicted phase curves of the planet, as shown in \citet{Tan_2019}. The resulting phase curves of their modeled planets had smaller offsets and larger amplitudes when the atmospheric drag was stronger.  This means that increasing the strength of the drag resulted in hotter daysides and cooler nightsides on the planet as well as reduced the eastward shift of the planet's hotspot, as should be expected. 

A limitation of universal drag timescales is that they do not allow for spatial variation in drag strength between the hot day and cold nightside of the planet: given the hundreds of Kelvin temperature contrasts, the magnetic resistivity should vary by many orders of magnitude, \deleted{and} so we should expect effective magnetic timescales to also vary by orders of magnitude for a single pressure level \citep[][]{RauscherMenou2013}. Uniform drag timescales also do not account for the directional dependence of magnetic drag: assuming the planet's deep-seated magnetic field is a dipole, only winds in the east-west direction should experience \added{Lorentz} drag\citep{Perna2010}. 

The most complex treatments of magnetism in the atmosphere of a hot gas giant involve the use of magnetohydrodynamic (MHD) models. \citet{Rogers_2014_showman} introduced the first \added{non-ideal} MHD simulation of a hot Jupiter. The authors compared the Lorentz force from their MHD simulations to the active magnetic drag timescale prescription from \citet{RauscherMenou2013}. Their Lorentz force peak value was within an order of magnitude of the timescale prescription, but the spatial extent of their Lorentz force was more localized. In a follow up paper, \citet{Rogers_2014b} ran a grid of MHD hot Jupiter models and found that time variability becomes important at the higher temperatures modeled (1400-1800 K) pointing to the intrinsic importance of feedback between magnetism and the planet's thermal structure. In the UHJ regime, \citet{Rogers2017} offers the closest instance of a physically consistent non-ideal magnetohydrodynamic treatment of an UHJ.  This work, focused on HAT-P-7b, highlighted the complexity of dynamic magnetic field lines and estimated the minimum global magnetic field strength of the planet to be 6 Gauss, based on comparison to observed phase curve variability for this planet \citep[][]{Armstrong2016}. This minimum field strength of 6 Gauss is also inferred from shallow-water magnetohydrodynamic models from \citet{Hindle2019}.

Magnetism is not the only high-temperature physical process of note in these planets. If high enough temperatures are reached on the dayside of UHJs, H$_{2}$ is expected to dissociate,  which results in a local cooling effect as dissociation requires an input of energy.  As winds transport the gas to the nightside of the planet, the temperatures can drop enough that recombination occurs, locally heating the atmosphere and reducing the day-night temperature contrast \citep{Bell_2018}. In \cite{Tan_2019}, the authors investigated the effects of molecular hydrogen dissociation and recombination, in addition to the presence of uniform drag, across a range of planetary equilibrium temperatures.  Their work found that beginning near $T_{eq}=2200$ K, the dissociation and recombination of hydrogen begins to play a significant role in the circulation of the atmosphere. Including hydrogen dissociation decreased the strength of the equatorial jet, which disappeared at the highest equilibrium temperature examined ($T_{eq}=3600$). Overall, including molecular hydrogen dissociation and recombination reduced the day-night temperature contrast, reducing the predicted phase curve amplitude. 

Beyond H$_2$, there are other molecular species whose dissociation has significant impact on the atmospheric structure of UHJs by changing the main sources of opacity. \citet{Parmentier2018} describes the influence of thermal dissociation and ionization of a variety of species including H$_{2}$O and H$^-$ on the thermal and spectral properties of UHJs, using four UHJ models made from SPARC/MITgcm. (The only main molecular opacity source that did not dissociate anywhere in the UHJ atmospheres was CO.)  These models showed stark differences between day and nightside temperatures and abundances. \added{They found} \deleted{The} dissociation of water and presence of H$^-$ opacity may result in muted of spectral features. Additionally, when included as an opacity source,  H$^-$  increases the optical depth across all wavelengths, causing the photosphere to move to a lower pressure \citep[][]{Lothringer2018}.  

Many UHJs exhibit what are known as temperature inversions---portions of the atmosphere that increase in temperature with decreasing pressure, opposite to standard behavior. These hot upper atmosphere \added{inversions} are likely due to high altitude optical/UV absorbers, such as TiO/VO \citep{Hubney2003, Fortney2008} or Fe, SiO, and metal hydrides \citep{Lothringer2018,Gibson2020}, but high C/O ratios \citep{Molliere2015} or inefficient IR cooling \citep[][]{Gandhi2019} could also play a role. 
The strength of this inversion may also scale with stellar host type \citep{Baxter2020}. Additionally, temperature inversions on UHJs are spatially inhomogeneous, as the stratospheric heating on the dayside will be absent from the shadowed nightside \citep{Kreidberg_2018}.

Due to the high temperatures of the daysides of UHJs, clouds are expected to be mostly constrained to the nightside of the planets.
Recently, \cite{Roman_2021} explored cloud effects in our GCM for a variety of irradiation temperatures, including UHJs. This work found that in the UHJ regime ($T_{irr}>3250K$),\footnote{A planet's irradiation temperature is related to its (zero-albedo) equilibrium temperature as: $T_{\mathrm{irr}}=4^{1/4} T_{\mathrm{eq}}$.} clouds were restricted to the nightside at high latitudes and were absent near the equator globally. Using a more complex cloud microphysics model, the GCM from \citet{Mansfield_2018} was post-processed in \citet{Helling2019}, \replaced{where they found}{which concluded} that some clouds would exist on the nightside of the UHJ HAT-P-7b and the dayside could host some very optically thin clouds, away from the equator. Their similar analysis of WASP-18b in \citet{Helling2019Wasp18} also found a cloud-free dayside and heterogeneous clouds on the nightside. As far as hazes are concerned, \citet{Helling2020} showed that the dayside of all UHJs are too hot for hydrocarbon hazes to be stable and that these hazes should play no role in the aerosol opacities on the nightside and terminator region.\deleted{, for the case of WASP-43b.}  

One particularly interesting UHJ is WASP-76b. First detected by \citet{West2016}, WASP-76b is an inflated ultra-hot Jupiter orbiting every 1.81 days with an equilibrium temperature just over 2200 K.  
Since its detection, a broadened sodium feature was detected in transmission spectra by \citet{Seidel2019}, hinting at the super-rotation of the upper atmosphere. The sodium detection was confirmed shortly after in \citet{zak_2019}. \citet{Ehrenreich2020} detected Fe absorption in transmission spectra from one side of the planet's terminator but not the other, which was interpreted as evidence for nightside condensation of the species. This detection was later confirmed by \citet{Kesseli2021}. \added{Recent work from \citet{wardenier2021} suggests that this differential absorption could instead be explained by a strong temperature difference on the trailing and leading limb of the planet, without needing to invoke iron condensation. } \citet{Fu2020} additionally detected TiO and H$_{2}$O in transmission spectra form $HST$ and $Spitzer$. From the emission spectra of the planet, CO emission features are present and their models suggested a temperature inversion of $\sim 500 K$. HST emission and transmission spectra also suggests the presence of TiO, H$_{2}$O, and thermal inversions \citep{Edwards_2020}. Tentative detections ($4 \sigma$) of VO also exist  \citep{Tsiaras_2018}.

Perhaps due to their difficulty to model, (\deleted{because}\added{as a result} of their extremely short radiative and dynamic timescales, in addition to the complicating physics discussed above) only a handful of GCMs for UHJs have been published. Because atmospheric dynamics will manifest in observables, 3D GCMs are  useful in interpreting the spectra, \citep[such as HAT-P-7b in][]{Mansfield_2018}  phase curves, \citep[such as the case of WASP-103b in][]{Kreidberg_2018} or both \citep[see WASP-18b from][]{Arcangeli2019}. A common theme that arises from these works is that the GCM had difficulty reproducing  UHJs with very low heat redistribution, which could potentially be due to an underestimation of magnetic drag strength, uncertainties surrounding dissociation effects, or warming from nightside clouds. When additional sources of drag are included, the resulting phase curves produced by the GCM in the works referenced above are a better match to observed values. 

In this work, we explore the effects of active magnetic drag in our GCM of the UHJ, WASP-76b. Although all of the physical processes mentioned above are no doubt present to some degree on WASP-76b, we choose to not include clouds or H$_2$ dissociation effects at this time to focus solely on the influence of active magnetic drag in the planet's atmosphere. Future work is necessary to characterize the mutual interactions between these physical processes for UHJs. Based on observational constraints on the strength of the temperature inversion in this planet's atmosphere \citep{Fu2020}, we set up our model to include this stratosphere. In Section \ref{Methods}, we describe our GCM and the active magnetic drag treatment. In Section \ref{Results}, we examine the atmospheric structures of our models with and without the active magnetic drag treatment. We additionally compute models featuring a universal drag timescale and compare the atmospheric effects of the two different treatments of magnetic drag. In Section \ref{Discussion}, we contextualize our results and discuss the limitations of our models. Finally, in Section \ref{Concl} we summarize the main points of this work.

\section{Methods} \label{Methods}
General Circulation Models (GCMs) are three-dimensional numerical tools  that  simulate the underlying physics and circulation patterns of planetary atmospheres. To do this, GCMs  solve the simplified set of fluid dynamics equations known as the primitive equations of meteorology. We use the GCM from \cite{GCM} with the updated radiative transfer scheme in \cite{newradRomanRausher} \citep[based on][]{Toon1989}.  Our GCM solves the radiative transfer with a double-gray treatment, meaning that two absorption coefficients are used: one in the visible wavelength regime to account for absorption from the host star and one in the infrared regime for the planet's thermal emission\added{The nuances of atmospheric implications for doubley-gray radiative transfer versus more complex treatments, such as correlated-k or picket fence methods, are beyond the scope of this work. Interested readers are directed to the recent paper from \citet{Lee2021} that explores in depth the consequences of these  different radiative transfer schemes.} We modeled the planet with 65 vertical layers evenly spaced in log pressure \added{over 7 orders of magnitude from the bottom boundary of 100 bars}, at a horizontal spectral T31 resolution (roughly 4 degrees at the equator) and the parameters listed in Table \ref{tab:gcm_params}. Our absorption coefficients were informed by \citet{Fu2020}; by changing the ratio of the infrared and visible coefficients, our resultant atmosphere will have an inverted temperature profile at locations that are highly irradiated.

Due to the increased difficulty of modeling planets in the ultra-hot regime, where heating rates are stronger and winds can be faster, we added sponge layers to the top three layers of all the models presented for numerical stability purposes.  Sponge layers act as a damping mechanism in the top of atmospheric models to reduce the buildup of artificial noise brought about by atmospheric waves reflecting off of the top boundary of the model \citep[][]{Forget1999,Wills2016}. The strength of these sponge layers decreases with height, with the strongest drag being applied in the top layer and then linearly decreasing with log pressure. By restricting the sponge layers to the top few levels, we leave the rest of the atmosphere unaffected. In our initial investigations, we found that after 1000 orbits were completed, the overall kinetic energy in the sponge layers was decreased by a factor of a few percent, and below the sponge layers, the change was even smaller. This is the first time sponge layers have been used in our GCM, but sponge layers are relatively common and have been implemented in other exoplanet GCMs  \citep[see][for examples]{Mayne2014, Deitrick_2020THOR, Wang_2020}. 

\begin{deluxetable}{lc}
\caption{WASP-76b Model Parameters} 
\label{tab:gcm_params}
\tablehead{ \colhead{Parameter} & \colhead{Value}} 
\startdata
         Planet radius, $R_{p}$ & $1.31 \times 10^{8}$ m \\
         Gravitational acceleration, $g$ & 6.825 m s$^{-2}$ \\
         \added{Orbital Period} & \added{1.81 days} \\
         Orbital revolution rate, $\omega_{\mathrm{orb}}$ & $4.018 \times 10^{-5}$ s$^{-1}$ \\
         Substellar irradiation, $F_{\mathrm{irr}}$ & $5.14 \times 10^{6}$ W m$^{-2}$\\
         Planet internal heat flux, $F_{\mathrm{int}}$ & 3500 W m$^{-2}$\\
         Optical absorption coefficient, $\kappa_{vis}$ & $2.4 \times 10^{-2}$ cm$^{2}$ g$^{-1}$ \\
         Infrared absorption coefficient, $\kappa_{IR}$ & $1 \times 10^{-2}$ cm $^{2}$ g$^{-1}$ \\
         Specific gas constant, $R$ & 3523 J kg$^{-1}$ K$^{-1} $\\
         Ratio of gas constant to heat capacity, $R/c_{p}$ & 0.286 \\
\enddata

\end{deluxetable}

\subsection{Our Magnetic Drag Treatment} 
\label{sec:mdrag}
\begin{figure*}
    \centering
    \includegraphics[width=.85\textwidth]{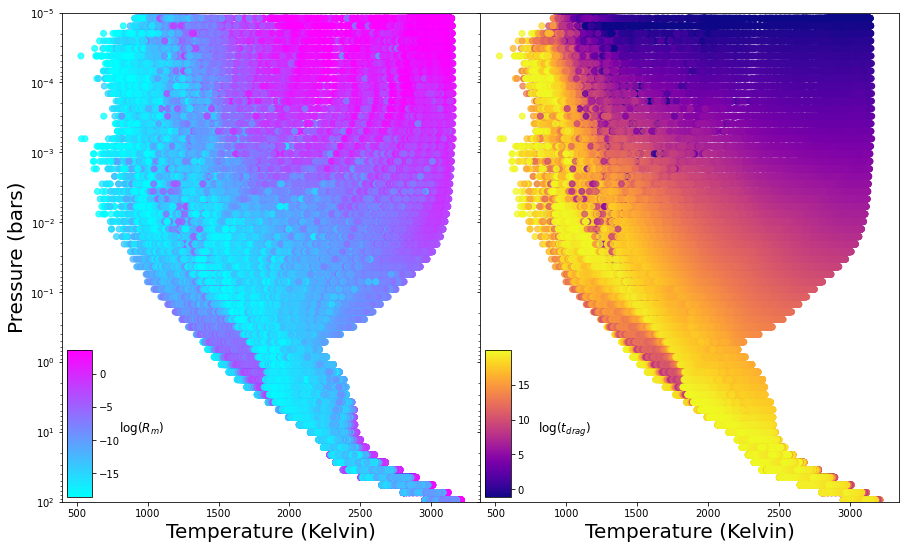}
    \caption{Magnetic Reynolds number for the 3 G model of WASP-76b \added{\textit{(left)} and the magnetic drag timescales \textit{(right)}} \added{based on the atmospheric structure at the end our simulation}. \added{ Our kinematic MHD framework is appropriate where $R_{m} < 1$, which is the case for the majority of the atmosphere. The hottest regions on the dayside upper atmosphere, where $R_{m}$ is the largest, is where atmospheric circulation can induce a magnetic field component comparable to the global field strength.  On the right, we show the range in values for our active drag timescale.  } Across a single pressure level, the magnetic drag timescale can vary by many orders of magnitude. Since our magnetic drag varies with latitude, a single temperature and pressure will have a range of timescales which explains the  non-monotonic behavior in portions of the plot.  } 
    \label{fig: tdrag3g}
\end{figure*}

\replaced{Although it does not solve the full non-ideal MHD equations,}{Our} model's implementation of magnetic drag is unique among GCMs as it is calculated based on local atmospheric conditions and applied in a geometrically and energetically consistent way. The physical origin of this drag comes from thermally ionized particles interacting with the planet's magnetic field due to strong (mostly neutral) winds advecting the particles across the planet. From the atmosphere's perspective, the wind feels a bulk Lorentz drag force. We choose to model the case of a dipole field aligned with the planet's rotation axis as a simplifying assumption,  because detailed information about the shape of exoplanet's magnetic field lines is not currently known. Additionally, we fix the shape of the magnetic field lines and \added{assume }the \added{local} field strength \added{does not vary as a function of radius} in the model. Our prescription for magnetism stands out from other GCMs because we use an active magnetic drag timescale; that is, one that is locally calculated and updated throughout the simulation to model the effects of magnetism in the planet's atmosphere. \added{We can refer to this framework as a ``kinematic'' MHD framework. The critical assumption with this treatment is that we are assuming the dipolar magnetic field, generated in the planet's interior, is much stronger than any field induced in the atmosphere. As a result of this assumption, we apply the drag only in the zonal (east-west) direction. This assumption remains reasonable when the magnetic Reynolds number is less than 1 \citep[][]{Menou_2012, Hindle2021a,Hindle2021b}. The magnetic Reynolds number can be approximated with $R_m \approx \frac{U H}{\eta}$\citep[][]{Hindle2021b} where $U$ is the zonal wind speed, $H$ is the pressure scale height, and $\eta$ is the magnetic resistivity. In this work, we calculate resistivity in the same way as \citet{Menou_2012}: 
\begin{equation} \label{resistivity}
    \eta = 230 \sqrt{T} / x_{e} \textnormal{ cm$^{2}$ s$^{-1}$}.
\end{equation}
We calculate the ionization fraction, $x_{e}$, using the Saha equation, taking into account the first ionization potential of all elements from hydrogen to nickel \citep[as in][]{RauscherMenou2013}.
As shown on the \replaced{right}{left} in Figure \ref{fig: tdrag3g}, $R_{m} <1$ for the majority of temperatures and pressures found in our model, implying that our assumptions are reasonable for much of the atmosphere, although we may be missing more complex behavior in the hottest dayside regions of the upper atmosphere.  } Although this is a simplification of \added{non-ideal} atmospheric MHD effects, it nevertheless represents a reasonable starting point for exploring the effects of active magnetic drag and is more complex than the uniform drag timescales applied in other GCMs.

\added{Our model also has unique advantages compared to current non-ideal MHD simulations in the low magnetic Reynolds number regime. In models such as those found in  \citet{Rogers_2014b}, magnetic resistivity is calculated from a static, horizontally homogenous (no latitude or longitude dependence) initial temperature profile. Other work, such as \citet{Rogers2017}, calculates resistivity based on a static temperature profile with horizontal variations, but does not change as the simulation runs and local temperature conditions change.  Our GCM calculates this resitivity locally and often, allowing the resitivity to evolve temporally in a more self consistent manner. By doing this, our resistivity varies by many orders of magnitude from the dayside to the nightside for a single pressure level and updates as the model progress, allowing for feedback with the atmospheric thermal structure.  Additionally, our model is coupled with double-gray radiative transfer equations. Current non-ideal MHD models lack this radiative transfer coupling and instead employ a  Newtonian relaxation for radiative forcing, a simplified scheme where the temperatures relax torward a prescribed profile at a chosen timescale \citep[][]{Rogers_2014_showman,Rogers_2014b}.  Finally, our model's dynamical core does not have any explicit viscosity in the momentum equations, as atmospheres are generally highly inviscid. Because current non-ideal MHD models such as those discussed above have heritage in dynamo/interior models with explicit viscosity in the solved equations, their resulting wind speeds are up to an order of magnitude weaker than expected from inviscid GCMs in the hot Jupiter regime. }

As described in detail in \cite{RauscherMenou2013} and shown in Equation \ref{tdrag}, the magnetic drag is applied \deleted{in an added drag term for the} \added{by subtracting $U/\tau_{mag}$ from the} east-west momentum equation, where the magnetic timescale is calculated locally as:
\begin{equation} \label{tdrag}
    \tau_{mag}(B,\rho,T, \phi) = \frac{4 \pi \rho \ \eta (\rho, T)}{B^{2} |sin(\phi) | }
\end{equation}
where $B$ is the chosen global magnetic field strength, $\phi$ is the latitude, $\rho$ is the density. \deleted{ and the magnetic resistivity ($\eta$) is calculated in the same way as \citet{Menou_2012}: 
\begin{equation} \label{resistivity}
    \eta = 230 \sqrt{T} / x_{e} \textnormal{ cm$^{2}$ s$^{-1}$}
\end{equation} }
 \added{The kinetic energy lost as a result of this drag is then returned to the atmosphere as localized ohmic heating in the energy equation as $\frac{U^{2}}{\tau_{mag}}$, for both the active drag discussed here and uniform drag models discussed below.  } This local treatment of magnetic drag allows the magnetic timescale to vary by many orders of magnitude throughout the entire planet without being as computationally expensive as solving the full non-ideal MHD equations. Additionally, this treatment updates the timescales as the model progresses, allowing feedback between the magnetic effects and the atmosphere's thermal structure. Figure \ref{fig: tdrag3g} \added{(right} shows the magnetic drag timescales across the atmosphere's temperature and pressure space from our 3 G model of WASP-76b. Across a single pressure level in the upper atmosphere, our magnetic drag timescale can vary by nearly 15 orders of magnitude.  The shortest magnetic timescales are located on the  dayside of the upper atmosphere, due to the high temperatures and low densities. Deeper in the atmosphere, where the density is greater, the timescales are longer and the effect of our magnetic drag on the circulation pattern will be smaller. This large variation in magnetic drag strength highlights the versatility of our active, locally calculated drag treatment. 

\added{Both $\tau_{mag}$ and $R_m^{-1}$ scale linearly with the magnetic resistivity and so in Figure \ref{fig: tdrag3g} we see some correlation between the two. However, the other physical variables that factor into the magnetic Reynolds number and drag timescale mean that they do not trace each other exactly. For example, almost all of the regions with $R_m>1$ also have $\tau_{mag}<10^5$ s, but there are locations with $\tau_{mag}=10^5$ s that have $R_m$ values as low as $10^{-10}$. Even the shortest drag timescales, $\tau_{mag} \sim 1$ s, can exist in regions with $R_m \sim 10^{-2}$, although generally those timescales are found in regions with $R_m \ge 1$.}

We ran our model of WASP-76b for variety of different magnetic field strengths: 0 G, 0.3 G, 3 G, and 30 G to encompass the field strengths measured in our solar system and to explore the range of effects our magnetic treatment will have on our modeled atmospheres. We additionally ran two instances of a uniform drag with timescales equal to $10^{4}$ and $10^{7}$ seconds following the GCMs in \citet{Tan_2019}. All of these models were calculated for 2000 orbits, to allow the atmosphere enough time for winds to accelerate (starting from rest) and reach a steady state. 

\section{Results} \label{Results}

\subsection{The Base Case: No Magnetic Effects}
\begin{figure}
    \centering
    \includegraphics[width=3.25in]{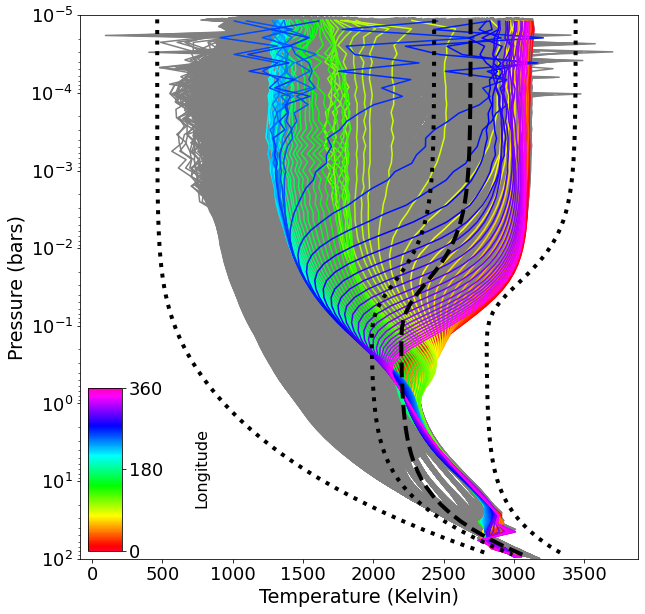}
    \caption{Temperature-pressure profiles for the non-magnetic model of WASP-76b. The rainbow lines show equatorial profiles (with longitude measured from the substellar point) while the gray curves show profiles from the entire planet. From right to left, the black dotted curves show the substellar, global, and nightside averaged \deleted{double-grey profiles.} \added{analytic 1D profiles calculated via \citet{guillot2010}.}  The dashed black line is the profile used to initialize conditions globally. A strong temperature inversion is present on the day side, which becomes non-existent on the un-illuminated nightside. The large day-to-night temperature contrast is also apparent.  } 
    \label{fig: tprof0g}
\end{figure}

To understand the effects of magnetic drag on our models of WASP-76b, we begin by examining the ``base" case atmosphere, free of magnetic drag \added{through a snapshot of the atmospheric structure taken from the end of the simulation}. Also, since this is the first planet with a forced inverted temperature profile we have published with this GCM, we will examine the extent and effects of the temperature inversion. Overall, our model of WASP-76b exhibits the typically expected hot Jupiter flow patterns seen in our previous works \citep[e.g.,][]{RauscherMenou2013,Beltz_2020,Roman_2021}. These features include a strong, eastward equatorial jet present in the deeper atmosphere (Figure \ref{fig: windscompare}) and an eastward advected hotspot. Figure \ref{fig: tprof0g}  shows the temperature-pressure profiles  for the planet as well as the 1-D profile we use globally to initialize the simulation and theoretical 1-D averages \citep[based on][]{guillot2010}.   From this figure, a few things are apparent. First, our model shows a  large ($>$1500 K) day-to-night temperature contrast along the equator. This is in broad agreement with the $\sim$ 1440K brightness temperature contrast measured from \textit{Spitzer} phase curves at 4.5 $\mu$m \citep[][]{May2021}. 

Second, while the temperature inversion must disappear on the nightside, as there is no stellar flux to be absorbed at high altitude, the east and west terminators are markedly different in the temperature inversions at those locations (roughly yellow and blue lines in Figure \ref{fig: tprof0g}, respectively). Because low pressures heat and cool rapidly,  the upper atmosphere temperatures are more similar at each terminator. The dominant eastward direction of the winds means that at deeper pressures the east terminator remains warm from gas heated on the dayside, while the west terminator has colder air advected around from the nightside.

Aside from the inverted temperature profile itself, another notable difference between this model and previous models with our GCM of non-inverted hot Jupiters is the decreased frequency and strength of chevron-shaped \added{temperature} features. These features are a result of \added{vertical motions caused by}converging winds, which heats the atmosphere in the chevron-shaped pattern \citep[e.g.,][]{Rauscher2010} and can can result in the hottest spot on particular pressure levels to be east \textit{or west} of the substellar point \citep[e.g.,][]{Beltz_2020}. In our model of WASP-76b however, these chevron features are weak enough that the hottest point stays east of the substellar point at all pressures (Figure \ref{fig: streammagplots}) and the chevron features are less prominent than in our non-inverted models.

\subsection{The Magnetic Models}
\begin{figure*}
    \centering
    \includegraphics[width=0.96\textwidth]{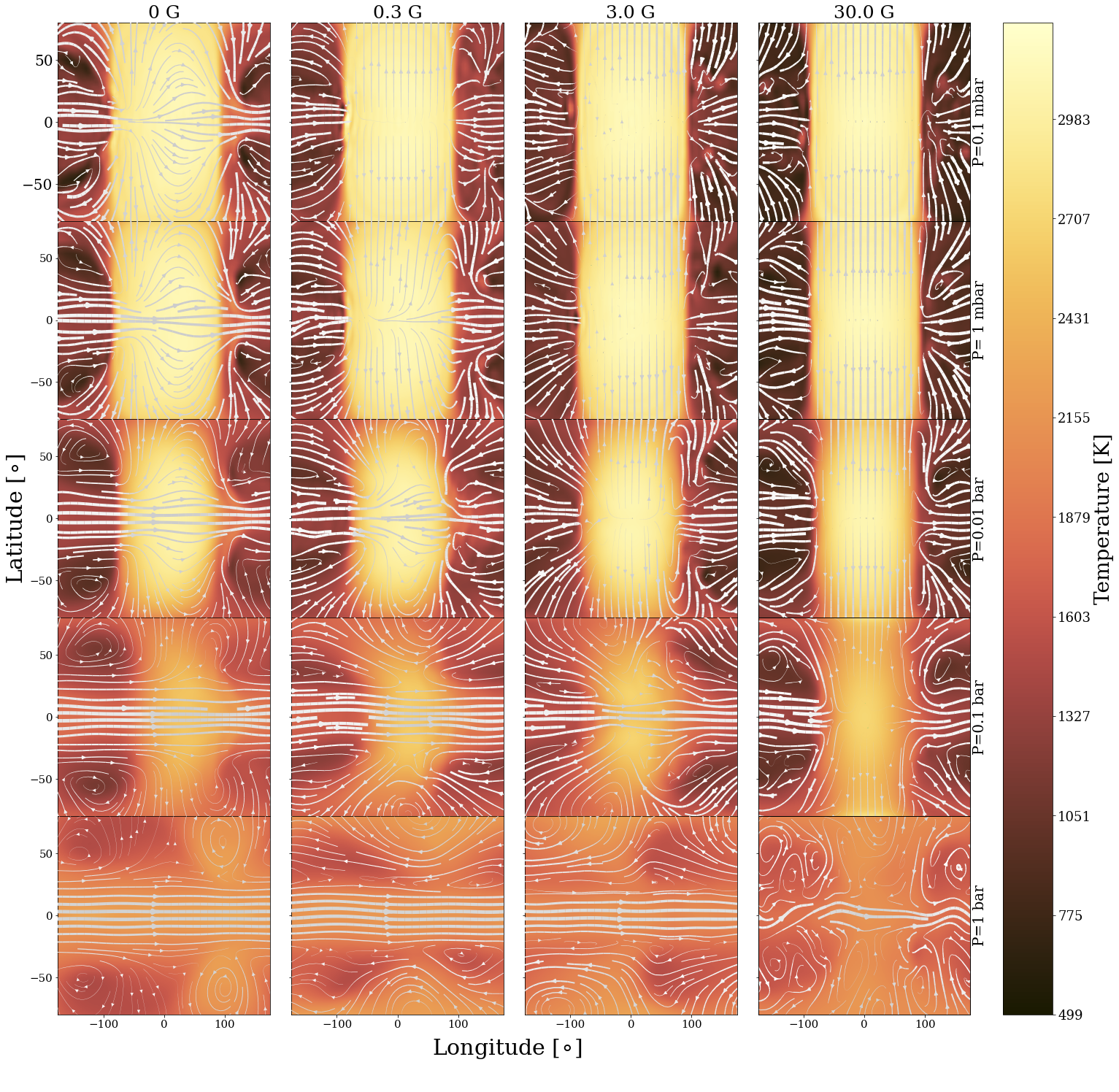}
    \caption{Maps of temperature and wind structure at several pressure levels (top to bottom: lowest to highest pressure) for each of the active magnetic drag cases studied (from left to right: increasing magnetic field strength). Each plot has the substellar point centered \added{and the thickness of the wind vectors scale with the relative wind speeds; maximum wind speeds for each map are reported in Table \ref{tab:maxwinds}}. The effects of our active magnetic drag are most readily seen at the lowest pressure, where all the magnetic cases show exclusively north-south flow on the dayside, as the magnetic treatment has removed all the east-west momentum.  As the pressure increases, the magnetic drag timescale also increases, reducing its effects. The stronger magnetic fields will have the shortest timescales and thus have the effects of magnetic drag present in deeper levels.  } 
    \label{fig: streammagplots}
\end{figure*}

We can now begin to look at the differences between our active magnetic drag models and our drag-free models. Figure \ref{fig: streammagplots} shows horizontal temperature maps for these scenarios at various depths in the atmosphere. The wind fields are overlaid in white stream lines. The differences between the non-magnetic models and magnetic models are most stark at the lowest pressures. The wind fields on the dayside of the magnetic models only flow in the north-south direction. This is because our magnetic drag is applied \textit{only} to the east-west momentum equation (see Section \ref{sec:mdrag}). At this low pressure and high temperature regime, the magnetic timescales are very short, even for the smallest magnetic field strength tested. This causes the elimination of east-west momentum, leaving only the weaker north-south wind vectors. As the pressure increases and we get deeper into the atmosphere, the choice of magnetic field strength results in more disparate atmospheric structures. At the second pressure level shown, P $ \sim 1$ mbar, the drag timescales become slightly longer, due to the increase in density. At this level, the weakest magnetic field, 0.3 G, shows some wind flowing in the east-west direction, but a majority of the flow remains in the north-south direction. The two strongest magnetic field cases on the other hand still show exclusively north-south flow. Going deeper into the atmosphere still, this pattern continues. At $P \sim 0.01 $ bar, the 0 G and 0.3 G cases \deleted{are nearly indistinguishable,} \added{are very similar} because the magnetic timescales at this pressure level have become so large that there is only a weak influence on the flow. \added{In the 0 G case, the equatorial jet is slightly stronger and broader. Additionally, the nightside vortices near $ \pm 50^{\circ} $ latitude are more pronounced than the 0.3 G model.}  The 3 G and 30 G cases still predominately have north-south winds on their dayside because the increased field strength allows the timescales to remain short enough to \added{strongly} affect the flow. At $P \sim 0.1$ bar, the magnetic timescales become so long that only the strongest magnetic field tested shows a \added{substantially} different flow pattern than the non-magnetic case. \added{At this pressure level, the 0 G case hosts the strongest equatorial jet, which becomes weaker as the magnetic field strength increases. } \deleted{Interestingly, at this pressure, the north-south flow causes a warming of the poles in the 30 G case.} \added{ Increasing the magnetic field strength from 3G to 30 G switches the atmospheric circulation pattern from a global equatorial super-rotation found in all models $<30$ G, to  a dayside dominated by day-night flow and nightside eastward equatorial flow. It is interesting that this eastward flow is maintained on the nightside, even though there is no longer eastward flow on the directly forced dayside. While we leave a detailed analysis of this flow pattern to future work, it is worth noting that such behavior is not found in models with uniform drag timescales.     } 

\begin{deluxetable}{cccccc}
\caption{Maximum  Wind Speeds (km/s) } 
\label{tab:maxwinds}
\tablehead{ \colhead{} & \colhead{0.1 mbar} & \colhead{1 mbar} & \colhead{0.01 bar} & \colhead{0.1 bar } & \colhead{1 bar}} 
\startdata
  \multicolumn{6}{c}{Active Drag Models}  \\ \hline
        0 G & 11.72 & 9.92 & 9.13 & 7.81 & 6.21 \\
        0.3 G & 12.44 & 8.96 & 6.66 & 5.65 & 4.30 \\
         3 G & \textit{7.21} & \textit{7.10} & 5.82 & 4.14 & 2.71 \\
         30 G & 6.52 & 4.89 & 4.47 & 2.72 & 0.86 \\ \hline
\multicolumn{6}{c}{Uniform Drag Models}  \\ \hline
         $10^{7}$s & 12.83 & 9.19 & 6.74 & 4.40 & 1.68 \\
         $10^{4}$s & 12.21 & \textit{7.97} & 4.62 & 2.09 & 0.24\\
\enddata
\tablenotetext{a}{Values in italics correspond to when the fastest winds are in the North-South direction. For all other models, the East-West winds are faster.}

\end{deluxetable}

\begin{figure*}
    \centering
    \includegraphics[width=20cm, height=8cm]{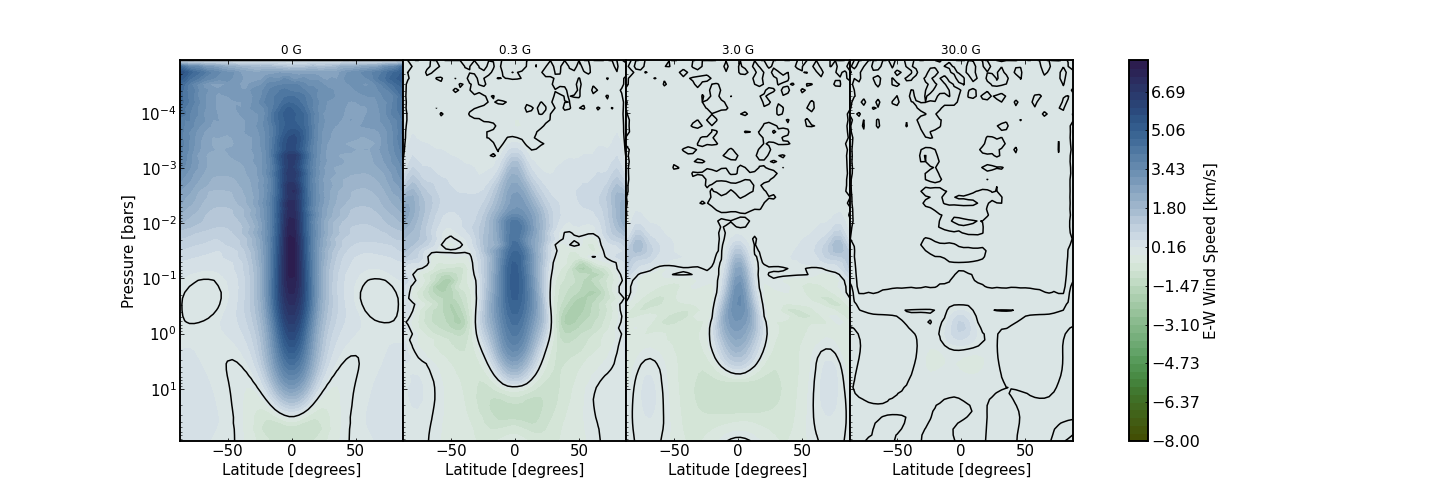}
    \caption{Longitudinally averaged east-west winds for the active magnetic drag cases studied. The black contours denote the boundary between positive (eastward) and negative (westward) winds. When the active magnetic drag is incorporated, the jet strength immediately diminishes, especially in the upper atmosphere. The stronger the magnetic field strength, the weaker the jet and the more it is constrained only to deeper pressure levels.  } 
    \label{fig: windscompare}
\end{figure*}

The equatorial jet is also affected by the magnetic drag, as seen in Figure \ref{fig: windscompare}. The non-magnetic case has the strongest and deepest jet. When active drag is present, the jet is weakened, especially at low pressures. This can be explained by the low pressures hosting the shortest drag timescales, which inhibit east-west flow and instead directs day-to-night flow meridionally up and over the poles. The stronger the magnetic field, the weaker the corresponding jet. In the strongest magnetic field we tested, the jet \added{almost} completely disappears. The reduction of the equatorial jet will result in less of an eastward shift of the planet's hotspot, meaning that the resulting phase curves will experience a smaller peak offset. 

\begin{figure*}
    \centering
    \includegraphics[width=0.9\textwidth]{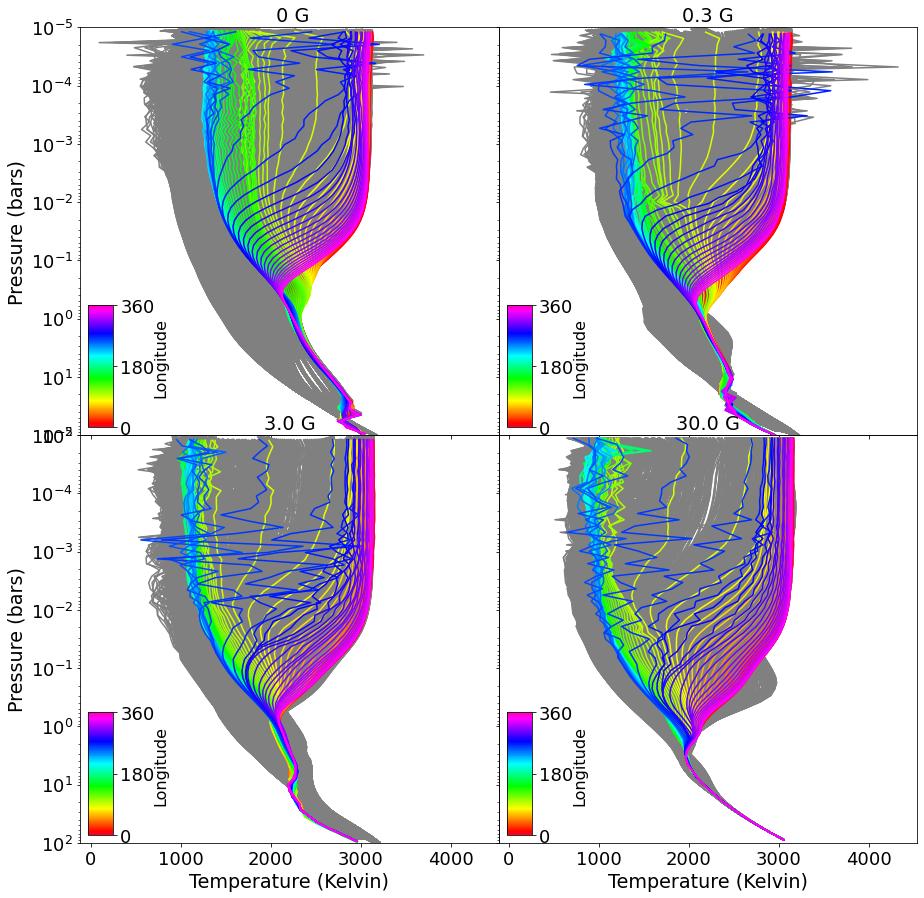}
    \caption{Temperature-Pressure profiles for our active magnetic models. The rainbow hues denote equatorial profiles, while the grey curves show profiles for the rest of the planet. All models show strong day-night contrasts and and temperature inversions on the daysides.  The increase in magnetic field strength causes the nightside temperatures to decrease and the lower atmospheres to become more uniform.} 
    \label{fig: tpactive}
\end{figure*}

In Figure \ref{fig: tpactive}, we show the temperature-pressure profiles for all of our active magnetic drag models. To first order, these profiles share many properties: the hottest parts of the dayside exceed 3000K, temperature inversions occur on the dayside but not the nightside, and the inversion varies in strength between the two extremes, most notably near the terminators. We can see the influence of the active magnetic drag in the nightside equatorial profiles;  as we increase the strength of the magnetic field, these profiles become colder. The models with magnetic drag also show more similarity between their east and west terminator profiles, in constrast to the non- or weakly-magnetic models where there is still significant eastward advection of hot gas on the dayside and cool gas from the nightside, leading to a warmer east terminator and cooler west side. Additionally, we see that at higher magnetic field strengths, the temperature profiles in the lower atmosphere become more uniform. 

We note that the increase in magnetic field strength also increases the amount of large temperature jumps between layers (zig-zag like behavior). This \added{may be} due to variability in heating and cooling at smaller spatial/temporal scales than our model resolves\added{could result from the numerical dissipation being too weak in these uppermost layers \citep[][]{Heng2011}, although since the best numerical dissipation strength to use should likely vary over the conditions throughout the atmosphere \citep[][]{Thrastarson2011}, increasing the dissipation in these upper layers would likely overdamp the higher pressure regions. } This variability is especially present with the dark blue curves, corresponding to near the terminator west of the substellar region. Given the fact that this region lies at the boundary between stellar irradiation and no irradiation, it is not surprising that these profiles exhibit these characteristics\added{, nor are they particularly worrisome.}  

\subsection{Uniform Timescale Drag Cases}
\begin{figure*}
    \centering
    \includegraphics[width=0.9\textwidth]{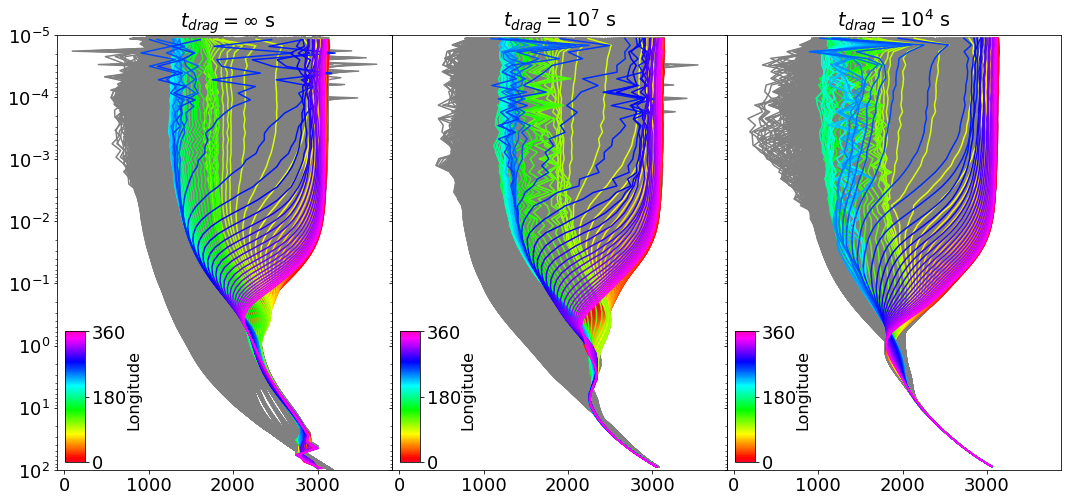}
    \caption{Temperature-Pressure profiles for different uniform drag timescales. The shortest uniform drag timescale case ($10^4$~s) exhibits a stronger thermal inversion ending deeper in the atmosphere and a more uniform lower atmosphere compared to the drag-free case. Additionally, the shortest uniform drag timescale causes cooler nightside temperatures, as it diminishes all of the temperature-homogenizing flow instead of just the east-west component, as in our active magnetic drag models. The $10^7$~s uniform drag timescale case shows these same effects, but to a lesser degree. } 
    \label{fig: universaldragtp}
\end{figure*}

We additionally wanted to compare the effects of our active magnetic drag treatment with  uniform drag timescale models. Following \citet{Tan_2019}, we ran our model of WASP-76b with uniform drag of timescales  $10^{4}$ and $10^{7}$ seconds. As in our active magnetic drag models, we included sponge layers for the top three layers in these models for numerical stability. 
The resulting temperature-pressure profiles for uniform drag are shown in Figure \ref{fig: universaldragtp}. Overall, these profiles are relatively similar to each other, especially in the upper atmosphere. The lower atmosphere (below $\sim 1$ bar) becomes increasingly uniform in temperature for shorter drag timescales. In fact, due to the uniform application of magnetic drag, the shortest drag model is essentially a single temperature at each pressure level below $\sim$10 bars. We saw a similar effect in our active magnetic models, but to a lesser degree. Notably, for the model with the shorter uniform drag timescale ($10^4~s$), the thermal inversion is slightly stronger and extends  deeper in the atmosphere. The longer uniform timescale ($10^7~s$) exhibits these effects to a smaller degree. This effect was not seen with our active magnetic drag models, where the timescales increase with increasing pressure. 

\begin{figure*}
    \centering
    \includegraphics[width=\textwidth]{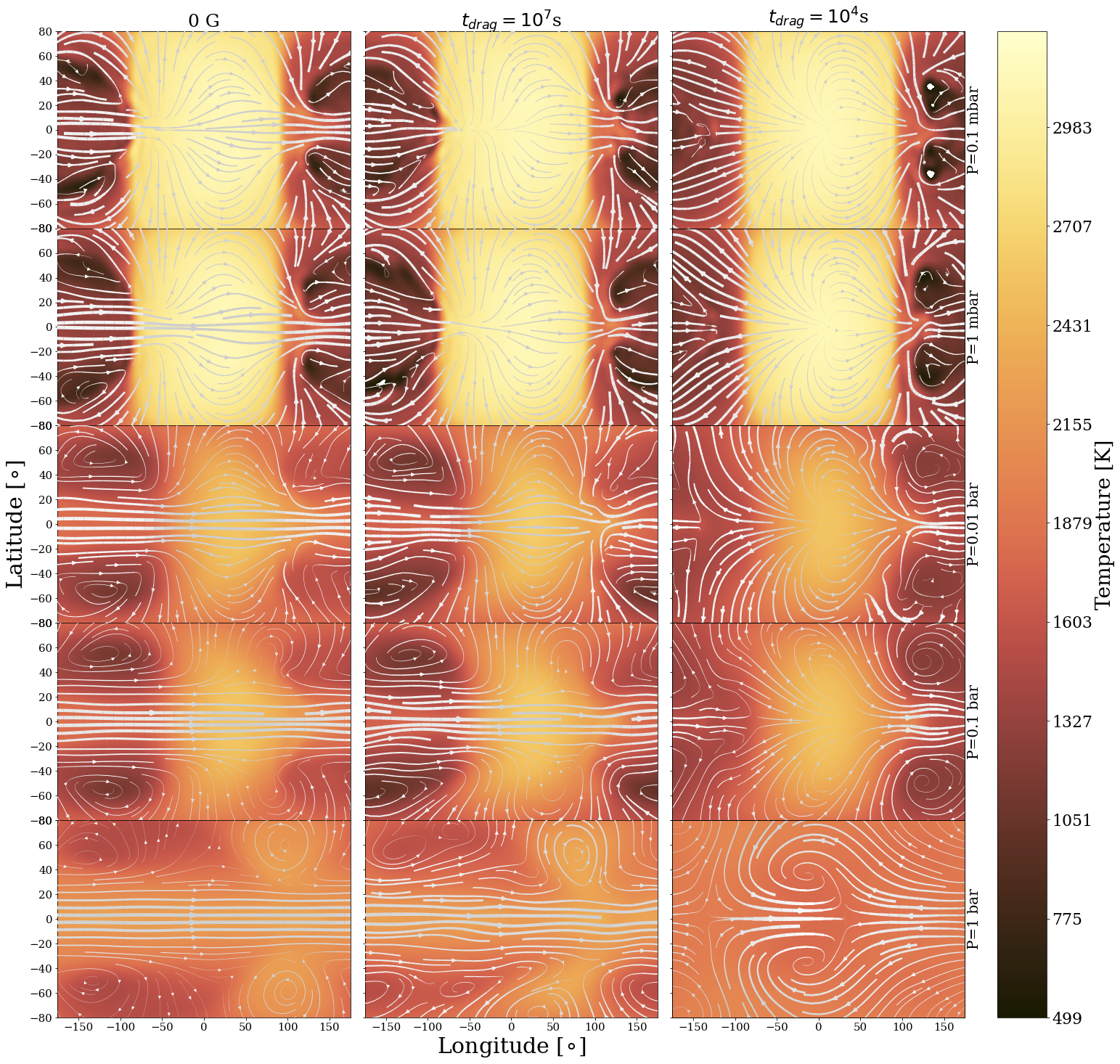}
    \caption{Maps of temperature and wind structure at the same pressure levels as Figure \ref{fig: streammagplots}, but comparing the uniform drag timescale cases. The differences in atmospheric circulation caused by the uniform drag timescale are more subtle but most prominently seen in the weakening of the equatorial jet in the lower atmosphere. } 
    \label{fig: streammagplotconsttdrag}
\end{figure*}

In Figure \ref{fig: streammagplotconsttdrag}, we show a similar plot to Figure \ref{fig: streammagplots}, but for the case of uniform drag timescale. The differences between the drag-free model and the uniform drag models are more subtle than for our active magnetic drag models with different assumed field strengths. Because the uniform drag is applied evenly throughout the atmosphere, the resulting wind structure is similar in shape but weaker in strength than the drag-free case \added{for the longer timescale. For the shorter timescale, the flow pattern west of the substellar point actually shows some westward motion at nearly all latitudes at and above 0.1 bar. This behavior is seen in other GCMs with uniform drag and is a result of the frictional drag becoming the dominant term in the momentum equation \citep[][]{KomacekShowman2016,May2021}. } \deleted{The greatest deviation from the drag free model is seen  in the P $\sim .01$ bar case for the short drag timescale, where the equatorial jet is significantly weakened. The longer constant drag timescale model also has a weakened jet, but not to the extent of the shorter drag timescale model.} 

\begin{figure*}
    \centering
    \includegraphics[width=15cm, height=8cm]{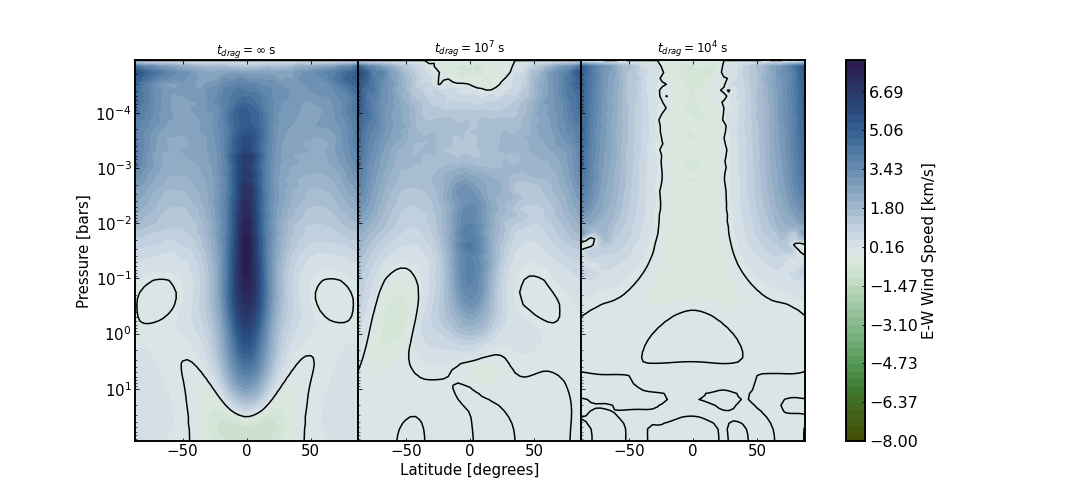}
    \caption{Longitudinally averaged east-west wind speeds for the uniform drag timescale cases. The $10^7$~s uniform drag timescale weakens the equatorial jet, while the shorter, $10^4$~s timescale completely eliminates it. In this shorter timescale we see instead weaker, higher latitude jets, which also begin to appear in the $10^7$~s drag timescale case.   } 
    \label{fig: windsconsttdrag}
\end{figure*}

Figure \ref{fig: windsconsttdrag} shows the longitudinally averaged east-west winds (akin to Figure \ref{fig: windscompare}) for the uniform drag timescale cases. We can see that the uniform drag timescale prescription destroyed the equatorial jet for the shortest, $10^4$~s timescale. Interestingly, some weaker, higher latitude eastward jets are present for most of atmosphere\replaced{above 1 bar}{($P \lesssim 1$ bar).} For the longer uniform drag timescale, we see that the jet has been weakened\replaced{, but it is still present throughout, except for the uppermost atmosphere where there are hints of the high-latitude jets}{to a point where there is actually a slight net westward flow at very low pressures ($P \lesssim 0.1$ mbar). However, from examining Figure \ref{fig: streammagplotconsttdrag}, we see that the wind structure contains both significant east and westward flow, largely canceling each other out in the zonal average. }. 

We can make a \deleted{apples-to-apples} \added{more direct} comparison of our uniform drag timescale models to the  ``cold interior'' models of WASP-76b in \citet{May2021}. For the weak drag timescale ($10^{7}$ s), the flow patterns and temperature structures between our two works (our Figure \ref{fig: streammagplotconsttdrag} and their Figure 5) are very similar. One difference to note is that our upper atmospheres tend to be hotter and our lower atmospheres are colder than those in \citet{May2021}, likely due to their inclusion of hydrogen dissociation and recombination. We also note that the longitudinally averaged east-west wind speeds (our Figure \ref{fig: windsconsttdrag} and their Figure 7) are \added{ broadly similar in overall wind pattern in the case of weak drag. Both models exhibit  a strong equatorial jet of similar strength that extends to $\sim 1$ bar. The model in \citet{May2021} exhibits westward flow at high latitudes on order of a few km/s that are not present in our model. Comparing the strong drag cases shows larger differences between our models. Although both models display the disruption of the equatorial jet at all pressure levels modeled, our model shows the emergence of net eastward flow at high latitudes extending to the top of the modeled atmosphere. The corresponding models in \citet{May2021} show westward flow for high latitudes at the top of the atmosphere ($P < 10^{-2} $bars) and eastward flow of a smaller magnitude at pressures lower than this., 
The general qualitative agreement between these uniform drag models (which differ in other parameter choices) suggests that the differences we find between our active and uniform drag models may be contextualized within the broader work of the community, where uniform drag timescales have been more commonly used.} \deleted{in good agreement. One slight difference is that for our strong drag timescale ($10^{4}$ s), the high latitude jets that form are stronger, though not as strong as those shown in their strongest drag timescale of $10^{3}$ s. This difference is minor and both of our works show the disruption of the equatorial jet at this timescale. }     

\subsection{Observable Consequences: Phase Curves}

\begin{figure*}
    \centering
    \includegraphics[width=0.9\textwidth]{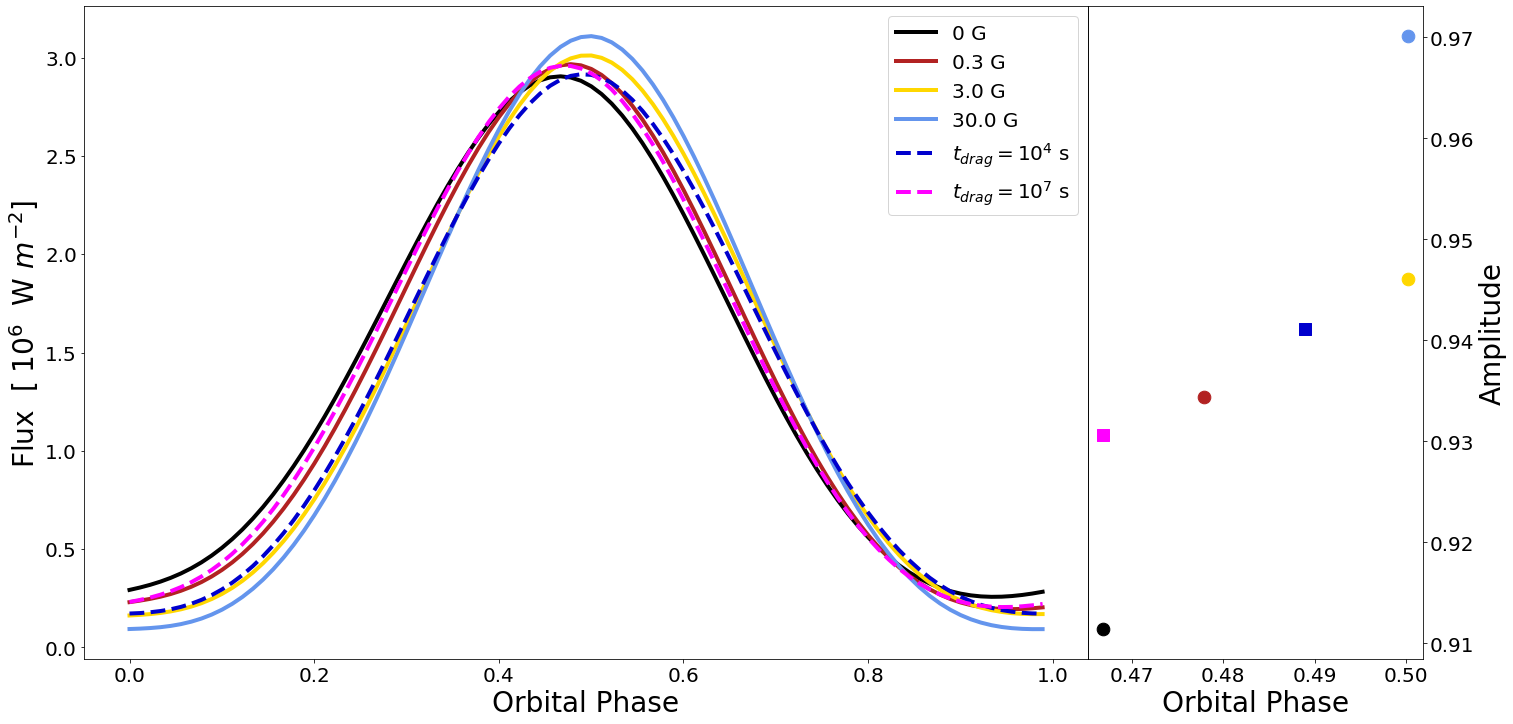}
    \caption{Left: orbital phase curves of total (bolometric) thermal emission for all of the models from this paper. Right: the amplitude ($(F_{\mathrm{max}}-F_{\mathrm{min}})/F_{\mathrm{max}}$) and phase of peak flux for each model's phase curve. For our active magnetic drag cases, increasing the magnetic field strength increases the amplitude of the phase curve and causes it to peak closer to a phase of 0.5 (which is secondary eclipse, not shown). This reflects the diminished day-night heat transport and disrupted eastward flow, due to the influence of the drag. The uniform drag timescale cases show similar behavior. The longer uniform drag timescale case has a similar hot spot shift to our non-dragged, 0 G case but an amplitude more similar to our 0.3 G model, while the shorter drag timescale has a phase curve amplitude and peak between the 0.3 and 3 G active drag models.  } 
    \label{fig: phasecurves}
\end{figure*}

Figure \ref{fig: phasecurves} shows our calculated phase curves from all of the models. As we increase the magnetic field strength in our active magnetic drag models, the peak of the phase curve shifts nearer to 0.5 (secondary eclipse) and the amplitude is increased. This is the expected result of our magnetic drag disrupting the eastward advection of hot gas away from the substellar point and generally reducing the transport of heat to the nightside. The difference in the phase of peak flux between our 3 G and 30 G models is small, as in both cases the hot spot is essentially unmoved from the substellar point. The larger amplitude of the 30 G case can be attributed to a cooler nightside and a stronger suppression of eastward flow in the upper atmosphere. A similar pattern exists for the uniform drag timescale models. The longer uniform drag timescale peaks at a similar phase to our model with no magnetic (or other) drag applied, meaning that its hot spot is still efficiently advected east of the substellar point despite the drag applied.  The phase curve for the shorter uniform drag timescale model peaks between the curves for our 0.3 and 3 G models, reflective of the increased ability of the drag to prevent advection of the hot spot.  Because phase curves integrate flux over the entire visible hemisphere, many of the differences between our active magnetic cases and uniform are lost in this integration.  Published Spitzer 4.5 micron phase curves of this planet show a very small hotspot offset ($<0.003$ in orbital phase from 0.5, \citet{May2021}), making our 3 G and 30 G models most consistent with the data. Given the double-grey nature of our GCM, comparing our bolometric phase curve amplitudes and the \textit{Spitzer} amplitudes in parts per million of the stellar flux is not trivial and imprecise, as it relies on the use of multiple assumptions. Thus, we refrain from those comparisons here.  \added{We do note however that our equatorial day-night contrasts are in broad agreement with the $~1440 $K brightness temperature contrast reported in \citet{May2021}.  }

\section{Discussion} \label{Discussion}

Compared to the less complex treatment of magnetism as a uniform drag, our active magnetic drag has some advantages. First, we are able to see clearly the difference in the magnitude of magnetic effects on the dayside vs the nightside. The uniform drag scenario forces all sides of the planet to be treated equally, despite the large difference in temperature around the planet. Many of the small scale atmospheric effects of our active magnetic drag disappear when the drag becomes uniformly applied.  Similarly to \citet{Kreidberg_2018}, we can roughly estimate the magnetic field strength that corresponds to  uniform drag timescales used in this work. Because the same timescale is applied to the atmosphere regardless of temperature, the corresponding field strength on the dayside and the nightside will differ dramatically. Based on equation 12 from \citet{Perna2010}, we estimate that our shorter drag timescale ($10^4$ s) would correspond to a field strength of $\sim$6 G on the dayside equator and nearly 300 G on the nightside equator.\footnote{For this calculation we used a pressure of  0.11 bar and temperatures of 2500K on the dayside and 1500K on the nightside. }  For the longer uniform drag timescale ($10^7$ s) we estimate the corresponding field strengths to be $\sim$0.2 G and $\sim$10 G on the dayside and nightside respectively. Physically, we do not expect the nightside field strength to be nearly two orders of magnitude stronger than the dayside. \added{Allowing our magnetic resistivity to vary spatially and temporally is also an improvement to treatments found in current non-ideal MHD models which employ temporally fixed resistivity. } Because our active magnetic drag treatment allows the drag strength to vary as a function of temperature and pressure, we are able to model magnetic \deleted{effects} \added{resistivity} in a more physically \added{consistent} \deleted{motivated} fashion.

Second, it is important that our active magnetic drag is only applied to the east-west momentum equation as we are assuming a planetary dipole field. As a result, there is a fundamental change in the direction of atmospheric circulation. Our dayside flow patterns were very distinct from the 0 G model in the upper atmosphere, as shown in Figure \ref{fig: streammagplots}. Uniform drag timescales, on the other hand, are applied to east-west \textit{and} north-south momentum equations. \deleted{As a result,} \added{When the uniform drag timescale is long enough that equatorial super-rotation is still present, }the corresponding flow patterns are similar in shape to the drag-free models but lower in strength. \added{The shorter uniform drag model was able to disrupt this superrotation in a way that can produce slight net westward flow near the equator. } Our active magnetic drag prescription is able to produce distinct differences in flow patterns that a uniform drag timescale is not. 

\added{ Many hot and ultra-hot Jupiters have inflated radii, which is likely due to energy deposited in the internal adiabat of the planet \citep[][]{Batygin2011,Laughlin2011,Lopez2016, KomacekYoudin2017,Thorngren2021}. Energy dissipation due to currents flowing in a resistive atmosphere has been theorized as a potential mechanism to deposit this energy \citep[][]{BatyginStevenson2010,Perna2010b,Rogers_2014_showman,Thorngren2018}.

In our models, we return the kinetic energy lost from our magnetic drag timescales in the energy equation as $u^{2}/t_{mag}$, mimicking local Ohmic dissipation. Notably, the deepest pressure modeled in this work is only 100 bars, so we cannot self-consistently predict the amount of heating that would be deposited deeper. However, using results from 1D global models of Ohmic dissipation profiles \citep[][]{Huang2012,Wu2013} as guidance, we might expect the Ohmic dissipation in the convective interior to be around $1 \% $ that of the Ohmic dissipation at the base of our dynamically active atmosphere. For each of global fields tested (0.3 G, 3.0 G, and 30 G) this approximates to $10^{18}$ W, $10^{20}$ W, amd $10^{19}$ W respectively deposited in the interior, which correspond to a maximum value of $\sim 0.01\%$ of the stellar irradiation for the 3G model. The internal heating needed to inflate a planet will be dependent on model assumptions and evolutionary history, but using the general estimate from \citet{KomacekYoudin2017}, which suggests that internal heating rates  $\lesssim 1\%$ of stellar irradiation  at a minimum depth of 100 bars can explain inflated hot Jupiter radii, we cannot conclude that our model's Ohmic heating could explain the inflated radius of WASP-76b.  Future studies, particularly those that model deeper pressures than the ones presented here, will be needed to explore this issue more thoroughly and self-consistently.   }

In order to first focus on the effects of our active magnetic drag treatment, we have chosen to omit molecular hydrogen dissociation/recombination, thermal ionization, and clouds. All of these processes should be present on WASP-76b, so further work is necessary to better understand the interactions of these processes with our active magnetic drag. Although we did not include these effects in our models, we can estimate how they would alter our model based on results from other GCMs. One of the main effects expected from molecular hydrogen dissociation and recombination is a reduction of the day/night temperature contrast \citep{Bell_2018,Tan_2019}. Cooling the dayside and warming the nightside would likely work to reduce some of the effects of magnetism on the upper atmosphere daysides of our models. By increasing the length of our magnetic timescale on the dayside of the planet, we would expect the day flow pattern of only north-south winds to not persist as deep in the atmosphere. The pressure at which east-west flows emerge again on the dayside will likely still be determined to first order by the strength of the global magnetic field. 

Based on the condensation curves from \citet{Roman_2021}, clouds would likely be confined entirely to the nightside, particularly at upper latitudes of the planet due to the high temperature of the dayside  These clouds could work to blanket part of the nightside, trapping heat and warming the area by a couple hundred Kelvin below the cloud.  This would decrease the magnetic drag timescales on these nightside pressure regions, but seeing that these timescales are already many orders of magnitude longer than the dayside timescales, the effects of our magnetic drag would still be predominately on the dayside of the planet. Furthermore, above the clouds,  we can expect a similar amount of cooling, which would decrease the likelihood of the overall nightside flow patterns being significantly altered from the cloud-free models we present here. 

Though we did not include clouds in these models, we can use the temperature profiles from Figure \ref{fig: tprof0g} and the condensations curves from \citet{Roman_2021} to speculate what potential cloud species could exist on the nightside of the planet. Based on the temperature structure, KCl, ZnS, and MnS clouds are only possible at high latitudes in the upper atmosphere, \added{at pressures lower than} \deleted{above} $\sim 10^{-2}$ bars or so. The condensation curves of Cr$_{2}$O$_{3}$, SiO$_{2}$, VO, and Mg$_{2}$SiO$_{4}$ show that these species could exist as condensates on the nightside equator \added{at pressures near} \deleted{above} $\sim 10^{-1}$ bars  and at high latitudes for lower pressures. Ni, Fe, Ca$_{2}$SiO$_{4}$, CaTiO$_{3}$, and Al$_{2}$O$_{3}$ are also potentially present at nearly all pressures modeled, but would be confined to the nightside of the planet.

Our active magnetic drag treatment  makes simplifying assumptions regarding the shape of the planet's magnetic field (a dipole), its orientation (aligned with the rotation axis), and that it lacks any time variability. To more accurately represent the magnetic field of this planet, \added{one potential route forward would be } \deleted{we would need} to incorporate non-ideal MHD equations, which are extremely computationally expensive in a 3D atmosphere. Work done at this level of complexity highlights the expected variability and asymmetry of the induced magnetic field lines and strength \citep[][]{Rogers_2014b, Rogers2017McElwain}, 
\added{and westward advection of the hot spot \citep[][]{Rogers2017,Hindle2021a}However, it is important to note that non-ideal MHD has never been coupled to a dynamical solver without explicit viscosity. Models with explicit viscosity return wind speeds significantly reduced compared to those found in primitive equation GCMs like the one presented here, even in the purely hydrodynamic case
\citep[][]{Rogers_2014_showman,Rogers_2014b}. Additonally, current non-ideal MHD models do not currently incorporate the large day-night magnetic resistivity variations presented in our model. A useful path forward could involve improvements to our kinematic MHD model by relaxing our strict dipole assumption and accounting for the toroidal component of the field. This would increase the validity of our model in the regions where the magnetic Reynolds number exceeds 1.  }
\deleted{If we were to use this treatment, we would still find that the strongest effects of magnetic drag would be found on the dayside of the planet, likely in the upper atmosphere. The resulting flow patterns would differ as well and could potentially show westward advection of the hot spot for periods of time 
The exact differences in the flow pattern would be strongly influenced by the orientation and strengths of the magnetic field lines throughout the atmosphere.}
\section{Conclusion} \label{Concl}
In this work we presented GCMs of the ultra hot Jupiter WASP-76b, focusing on the role of magnetic drag in shaping this planet's atmospheric structure. We used the ``active magnetic drag" treatment from \citet{RauscherMenou2013}, which calculates a drag timescales based on local conditions throughout the atmosphere and updates these values as the simulation runs. Our main results are as follows:
\begin{itemize}
    \item The influence of our active magnetic drag was most strongly seen in the upper atmosphere dayside of the planet, where the drag timescales were the shortest. The winds here were solely in the north-south direction, such that hot gas from the dayside mainly flowed to the nightside over the poles, as our magnetic drag reduces momentum in the east-west direction.
    
    \item The stronger the global magnetic field strength, the deeper in the atmosphere the effects of our magnetic drag were seen. Our drag treatment can dramatically disrupt the equatorial jet seen in hot Jupiter models, reducing its strength and causing it to be confined to the lower atmosphere, where the magnetic timescales become longer. At the strongest case we examined, 30 G, the equatorial jet was \deleted{completely destroyed.} \added{reduced significantly and the atmosphere no longer displayed superrotation. } 
    
    \item  We calculated bolometric thermal emission phase curves from our GCMs and found that the models with active magnetic drag had larger amplitudes (due to less efficient day-night heat transport) and peaked nearer to secondary eclipse. In comparison to Spitzer phase curve observations of this planet \citep[][]{May2021}, our models would imply a magnetic field strength of at least 3 G for WASP-76b. We do note, however, that other important physical processes are missing from these models (hydrogen dissociation and clouds) and a full comparison with the data requires modeling all of these processes together.

    \item We also looked at the effects of using a uniform drag timescale as an estimate for magnetic drag. Unlike our active drag timescale, the uniform treatment applied the same drag timescale globally in the east-west and north-south momentum equations. The circulation patterns in the upper atmosphere of the uniform case differed significantly from our active magnetic drag treatment, especially on the dayside.The corresponding phase curves put the longer timescale ($10^{7}$ s) model between our drag-free and 0.3 G active model and the short timescale ($10^{4}$ s) between our 0.3 G and 3.0 G active models. Order of magnitude estimates based on local conditions of the dayside and nightside of the planet indicate field strengths from $\sim$6-300 G and $\sim$0.2-10 G for the short and long timescale respectively. We have shown that our active magnetic drag case does a better job of reproducing the expected physics of UHJs.  
    
\end{itemize}

Our active magnetic drag prescription is an improvement over the use of uniform drag timescales to model the interaction between partially thermally ionized winds and a deep-seated planetary magnetic field, in that it captures more of the physical behavior we expect. However, we re-emphasize that the physically correct solution will require complex and computationally expensive full non-ideal MHD treatments \added{in regions where the magnetic Reynolds number is greater than 1. In this regime, the induced atmospheric field becomes comparable or larger than the dipolar field. As a result of this feedback in atmospheric circulation, hot spot reversal and time-variability can occur \citep[][]{Rogers_2014b,Hindle2019,Hindle2021a} Future changes to our model should improve on our kinematic MHD framework and incorporate the effects of the toroidal field on meridional winds to replicate these steady state features in the high Reynolds number regime. }. Given the high observational interest for UHJs, our  community should invest in further development and use of those models \added{especially for UHJs with strong thermal ionization such as WASP-76b, HAT-P-7b, WASP-18b, and KELT-9b.}

With future work, we will more carefully examine the influence of active magnetic drag on various types of atmospheric characterization measurements, beyond just the bolometric phase curves shown here. In particular, high-resoultion spectroscopy may be a promising avenue for empirically constraining wind speed and drag mechanisms \citep[][]{KemptonRauscher2012,Flowers2019,Beltz_2020}.
  \section{Acknowledgements}
\added{We would like to thank the reviewer for their detailed and incredibly helpful comments which greatly improved the quality of this paper. This work was generously supported by the Heising-Simons Foundation.}
\bibliographystyle{aasjournal}
\bibliography{bib.bib}

\end{document}